\pdfoutput=1
\AtBeginDocument{%
  \providecommand\BibTeX{{%
    \normalfont B\kern-0.5em{\scshape i\kern-0.25em b}\kern-0.8em\TeX}}}

\documentclass[sigconf]{acmart}

\usepackage{alltt}
\usepackage{multirow}
\usepackage{graphicx}
\usepackage{color}
\usepackage[labelfont=bf,textfont={bf}]{caption}
\usepackage{subcaption}
\usepackage{pifont}
\usepackage{boxedminipage}
\usepackage{notoccite}
\usepackage{array}
\newcolumntype{C}[1]{>{\centering\let\newline\\\arraybackslash\hspace{0pt}}m{#1}}
\usepackage{enumerate}
\usepackage{booktabs}
\usepackage[framemethod=tikz]{mdframed}
\usepackage{lipsum}
\usetikzlibrary{shadows}
\usepackage{mathtools}
 \usepackage{paralist}
 \definecolor{light-gray}{gray}{0.80}
\usepackage{graphics}
\usepackage[shortlabels]{enumitem}
\usepackage{dblfloatfix}
\newmdenv[
  tikzsetting= {fill=light-gray},
  linewidth=1pt,
  roundcorner=0pt, 
  shadow=false
]{myshadowbox}
\usepackage{colortbl} 
\usepackage{blindtext, graphicx}
\usepackage{textcomp}
\usepackage{mathtools}
\usepackage[final]{listings}
\usepackage{picture}
\usepackage{relsize}
\usepackage{multicol}
\usepackage{soul}
\usepackage{enumitem}
\setitemize{noitemsep,topsep=0pt,parsep=0pt,partopsep=0pt,leftmargin=*}
\setenumerate{noitemsep,topsep=0pt,parsep=0pt,partopsep=0pt,leftmargin=*}
\usepackage{makecell}
\usepackage{bigstrut}

\definecolor{comment_color}{rgb}{0.5, 0, 1}
\definecolor{steel}{rgb}{0.1, 0.3, 0.5} 

\newcommand{\bi}{\begin{itemize}}
\newcommand{\ei}{\end{itemize}}
\usepackage{verbatim}
\usepackage{algorithm}
\usepackage{algorithmicx}
\usepackage{algpseudocode}
\usepackage[export]{adjustbox}
\renewcommand{\footnotesize}{\scriptsize}

\definecolor{LightCyan}{rgb}{0.88,1,1}
\definecolor{darkgray}{gray}{0.7}
\definecolor{Gray}{rgb}{0.88,1,1}
\definecolor{Gray}{gray}{0.85}
\definecolor{Blue}{RGB}{0,29,193}
\definecolor{MyDarkBlue}{rgb}{0,0.08,0.45} 
\definecolor{pink}{rgb}{.96,.72,.77}
\definecolor{lightergray}{rgb}{0.85, 0.85, 0.85}
\definecolor{darkgray}{rgb}{0.47, 0.47, 0.47}
\definecolor{lightestgray}{rgb}{0.95, 0.95, 0.95}
\definecolor{ao(english)}{rgb}{0.0, 0.5, 0.0}
\definecolor{beige}{rgb}{0.96, 0.96, 0.86}
\usepackage{tcolorbox}
\newtcolorbox{blockquote}{colback=black!5,boxrule=0.4pt,colframe=black,fonttitle=\bfseries}
\newtcolorbox{RQbox}{colback=red!5,boxrule=0.4pt,colframe=black,top=1pt,bottom=1pt,fonttitle=\bfseries}

\copyrightyear{2020}
\acmYear{2020}
\setcopyright{acmcopyright}\acmConference[ASE '20]{35th IEEE/ACM International Conference on Automated Software Engineering}{September 21--25, 2020}{Virtual Event, Australia}
\acmBooktitle{35th IEEE/ACM International Conference on Automated Software Engineering (ASE '20), September 21--25, 2020, Virtual Event, Australia}
\acmPrice{15.00}
\acmDOI{10.1145/3324884.3418932}
\acmISBN{978-1-4503-6768-4/20/09}

\begin{document}

\title{  Making Fair ML Software using Trustworthy Explanation}

\author{Joymallya Chakraborty}
\email{jchakra@ncsu.edu}
\affiliation{%
  \institution{North Carolina State University}
}

\author{Kewen Peng}
\email{kpeng@ncsu.edu}
\affiliation{%
  \institution{North Carolina State University}
}

\author{Tim Menzies}
\email{timm@ieee.org}
\affiliation{%
  \institution{North Carolina State University}
}


\begin{abstract}

Machine learning software is being used in many applications (finance, hiring, admissions, criminal justice) having huge social impact. But sometimes the behavior of this software is biased and it shows discrimination based on some sensitive attributes such as sex, race etc. Prior works concentrated on finding and mitigating bias in ML models. A recent trend is using instance-based model-agnostic explanation methods such as LIME\cite{10.1145/2939672.2939778} to find out bias in the model prediction. Our work concentrates on finding shortcomings of current bias measures and explanation methods. We show how our proposed method based on K nearest neighbors can overcome those shortcomings and find the underlying bias of black box models. Our results are more trustworthy and helpful for the practitioners. Finally, We describe our future framework combining explanation and planning to build fair software. 
 
\end{abstract}

\maketitle

\section{Introduction}
\label{sec:intro}

There are many scenarios where machine learning software has been found to be biased and generating arguably unfair decisions. Sentiment analyzer model from Google which is used to determine positive or  negative  sentiment gives negative  score  to  some sentences like `I  am  a  Jew' and  `I  am  homosexual' \cite{Google_Sentiment}. Google translate shows gender bias, when ``She is an engineer, He is a nurse'' is translated into Turkish and then reverted back into English becomes ``He is an engineer, She is a nurse'' \cite{Caliskan183}. Amazon had to scrap an automated recruiting tool that became biased against women \cite{Amazon_Recruit}. 
A popular recidivism assessment model used by the criminal justice system shows racial discrimination\cite{Machine_Bias}. It predicts black defendants as future criminals with higher error rate than white defendants.

Researchers from Software Engineering and Machine Learning community have taken this social discrimination issue seriously and have started working on that. ACM and IEEE have started separate conferences like FAccT\cite{FAT}, FILA\cite{FILA} for \textit{fairness} of ML models. ASE 2019 organized EXPLAIN\cite{EXPLAIN} workshop where an important topic was \textit{ML software fairness}. Big industries such as IBM\cite{AIF360}, Microsoft\cite{FATE}, Facebook\cite{Fairness_Flow} have started putting efforts into this domain . IEEE\cite{IEEEethics} and European Union\cite{EU} published the ethical principles of AI. \textit{Fairness} has been given special importance there. It is stated that an intelligent system or machine learning software must be fair if it is used in real-life applications.

In the recent SE and ML literature, we see works in mainly three directions - testing ML software to find bias\cite{Galhotra_2017,Aggarwal:2019:BBF:3338906.3338937,Udeshi_2018}, mitigating bias in the model behavior\cite{NIPS2017_6988,zhang2018mitigating,chakraborty2020fairway} and using model-agnostic explanation methods to visualize bias or unfairness\cite{10.1145/2939672.2939778,NIPS2017_7062}. Machine learning fairness is a rapidly evolving research domain. It started gaining attention just ten years before but within this short time period, a huge number of papers have been published. Three years before, there were only five metrics to measure classification model bias where today there are more than seventy different metrics\cite{ClassificationMetric}. Kleinberg et al. stated that it is impossible to satisfy all the metrics simultaneously\cite{kleinberg2016inherent}. That means even if based on one metric the model looks fair but some other metrics may still complain. Berk et al. said that accuracy of a model and fairness are competing goals\cite{berk2017fairness}. These findings are increasing the ambiguity of ML software fairness and practitioners are getting confused about what to do; which metric to trust; which one to choose between a better predictor and a fair predictor.

 Due to these complications, a recent trend in fairness domain is to depend on interpretation or explanation tools. Two such extremely popular explanation tools are- LIME\cite{10.1145/2939672.2939778} and SHAP\cite{NIPS2017_7062} - they treat complex machine learning models as black-box and generate instance-based explanation in an interpretable manner. As these tools are very easy to use and a user can use these without having any knowledge about the machine learning model, practitioners have started using these to find the underlying bias of ML models\cite{AIFLIME}. However, some recent studies have found out these explanation tools are not trustworthy and should not be used to find bias in ML models\cite{Dimanov2020YouST,slack2019fooling}. The main reason is these explanation methods generate samples around the inspecting point by perturbing each feature. While doing that most of the samples generated become far from the original distribution of the training data. Thus, explanation based on those samples is completely misleading and not reliable. In this paper, we, instead of generating samples by random perturbation, use K-nearest neighbor approach to find similar data points from the training data. Our explanation approach is more trustworthy and makes it easier for domain experts to visualize the bias. This paper creates the buildings blocks of combining model fairness and explanation to create fair, trustworthy software. Overall, this paper makes the following contributions:

\bi

\item We comment on shortcomings of current fairness measures and explanation methods.
\item We propose a metric-free, nearest neighbor based approach to  overcome those shortcomings.
\item We evaluate our approach on a publicly available, widely used dataset.
\item Source code is publicly available on GitHub for future researchers\footnote{https://github.com/joymallyac/Fair-Knn}.
\item Finally, we describe our future directions towards generating policies and making plans to build fair \& trustworthy ML software.

\ei

\section{Background \& Prior works}
\label{Background}
 A ML software which is used for decision making is said to be unfair or biased if it gives undue advantages (being hired for a job, receiving loan) to a specific group of people (Sex-Male, Race-White) or individuals based on sensitive attributes. \textbf{Protected attribute} is an attribute that divides a population into two groups (privileged \& unprivileged) that have a difference in terms of benefits received. For example, in case of credit card application, ``Male'' is the privileged ``sex'' \& ``Female'' is the unprivileged. The goal of \textbf{Group Fairness} is based on the protected attribute, privileged and unprivileged groups will be treated similarly. According to IBM AIF360\cite{ClassificationMetric}, there are seventy different metrics to measure group fairness. Most of them mainly try to balance the True Positive Rate (TPR) and False Positive Rate (FPR) of both groups. Four of the most used group fairness metrics are mentioned here -

\bi
\item Demographic Parity (DP): The TPR \& FPR for both groups should be the same.
\item Equal Opportunity (EQ): The TPR for both groups should be the same.
\item Equal Accuracy (EA): The classifier accuracy for both groups should be the same.
\item Equal Odds (EO): The TPR and TNR for both groups should be the same.
\ei

\textbf{Individual Fairness} is the goal of similar individuals receiving similar outcomes. The similarity between individuals is application specific. Most of the time similar individuals are chosen based on some distance metric. The metric for measuring individual fairness is \textbf{Consistency}\cite{pmlr-v28-zemel13}.
\begin{equation}
   Consistency = 1 - 1/n.n_{neighbors} \sum_{i=1}^{n} | \hat{y_i} - \sum_{j \in N_{n_{neighbors(x_i)}}}^{n} \hat{y_j} |
\end{equation}

Prior works in ML software fairness are of mainly two types -

a> \textbf{Testing ML software to find bias} - Galhotra et al. created THEMIS\cite{Galhotra_2017}, a testing-based tool to find software discrimination, focusing on causality in discriminatory behavior. Udeshi et al. have developed AEQUITAS\cite{Udeshi_2018} tool that generates discriminatory inputs which find fairness violation. Aggarwal et al. have proposed a new testing method for black-box models\cite{Aggarwal:2019:BBF:3338906.3338937}. They combined dynamic symbolic execution and local explanation to generate test cases for non-interpretable models. These are  test case generation algorithms which find bias in the model behavior.

b> \textbf{Removing bias} - There are numerous algorithms to mitigate bias and achieve group fairness and individual fairness. In the case of group fairness, there are mainly three approaches - 

\bi
\item \textbf{Pre-processing algorithms}: Before classification, data is pre-processed in such a way that discrimination is reduced. Some popular works are - \textit{Reweighing}\cite{Kamiran2012}, \textit{Optimized pre-processing}\cite{NIPS2017_6988}.

\item \textbf{In-processing algorithms}: Here the dataset is divided into three sets - train, validation and test set. After model training, model is optimized on the validation set and finally applied on the test set. Some popular works are - \textit{Adversarial debiasing}  \cite{zhang2018mitigating}, \textit{Prejudice Remover} \cite{10.1007/978-3-642-33486-3_3}

\item \textbf{Post-processing algorithms}: Here some of the class labels are changed to reduce discrimination after classification. Some popular works are - \textit{Reject option classification} \cite{Kamiran:2018:ERO:3165328.3165686}, \textit{Equalized odds post-processing}\cite{pleiss2017fairness,hardt2016equality}.

\ei

The concept of individual fairness is ``similar individuals should be treated similarly''\cite{dwork2011fairness}. In most cases, similar individuals are chosen based on some distance metric such as Mahalanobis distance\cite{gillen2018online}. Some popular works to achieve individual fairness are - iFair\cite{lahoti2018ifair}, Metric-Fair Learning\cite{rothblum2018probably}, Learning Fair Representations\cite{pmlr-v28-zemel13}

\section{Fairness is hard to achieve}

In spite of so much prior works, machine learning fairness is still very ambiguous in nature and hard to achieve. Let's assume an interesting decision-making scenario. An employer is selecting candidates for interviews. They are using a ML model which uses application information and returns a prediction about whether a candidate would be a good employee or not. It is found that the model is more likely to give positive predictions for one gender (e.g. male) than others (e.g. female). Some Fair-ML technique is then used to prevent this. However, as a result of this bias mitigation, a male applicant is not selected for interview. He complains, pointing to examples of females who were selected despite having qualifications very similar to his. Now what should the employer do?  Should the employer continue to interview the female candidates or adjust the model again to ensure that any `more qualified' male get interview calls instead? So, there is conflict between Group Fairness \& Individual Fairness\cite{binns2019apparent}. If a model is optimized to achieve either of them, the other one may get damaged. Brun et al. mentioned that reducing bias for one attribute (e.g. sex) introduces bias for another attribute (e.g. race)\cite{FAIRWARE}. There are many metrics of fairness (e.g. 70 for group fairness\cite{ClassificationMetric}). But Berk et al. \& Kleinberg et al. mentioned that it is impossible to satisfy all kinds of fairness simultaneously\cite{Berk,kleinberg2016inherent}. Chakraborty et al. commented that current fairness metrics depend on absolute values of TPR \& FPR and completely ignore the possibility of imbalanced class distribution\cite{chakraborty2020fairway}. They raised questions about reliability of these metrics in real-world scenario. Thus fairness is very hard to quantify and hence harder to achieve because validation depends on metrics.

In attempt to solve these problems, we say that we should not just depend on some metrics or we should not just optimize ML models based on some metrics. If we can interpret \& visualize predicted outcomes generated by a model, then it will be easier for us to find the underlying bias and also rely on the findings.  LIME\cite{10.1145/2939672.2939778} \& SHAP\cite{NIPS2017_7062} are two recent extremely popular explanation tools. They provide instance-based explanation for any black box machine learning model. Both of them show the feature importance and feature contribution to the output for a single instance. The authors of these explanation tools claimed that these tools are trustworthy and generate reliable explanations. But recent findings show that these methods can be easily fooled\cite{slack2019fooling,Dimanov2020YouST}. Slack et al. used a scaffolding technique to make a fool of LIME \& SHAP and found that in some cases when a classifier is completely unfair (makes prediction based on only protected attribute) then also these explanation tools do not complain about model bias\cite{slack2019fooling}. Since the sample generator within LIME and SHAP presumes a normal distribution for the real data, usually it could generate out-of-distribution (OOD) synthetic data or samples. Thus explanations based on those samples can be not trustworthy and should not be used to find bias in model behavior. Dimanov \& other researchers \cite{Dimanov2020YouST} have reached similar conclusion that we can not trust these explanation methods to find the underlying bias of ML model. So, we think that explanation of model prediction is better way to visualize bias but current explanation methods are not good enough to be trusted. We need a trustworthy explanation strategy - that is where this paper contributes. 



\begin{table*}[]
\caption{The table shows five nearest data points (top 5 rows) for the row being inspected (last row) in the ``Adult'' dataset. Last row is predicted to have negative outcome by logistic regression model where first five rows have positive class label. ``capital-gain'' and ``capital-loss'' are the most important features which are same for all the rows. The main reason of the negative outcome for the last row is the value of protected attributes (race=Black and sex=Female) because that is the major difference from the nearest neighbors.}
\label{Adult}
\begin{tabular}{@{}cccccccccc@{}}
\toprule
\rowcolor[HTML]{C0C0C0} 
{\color[HTML]{000000} \textbf{age}} & {\color[HTML]{000000} \textbf{workclass}} & {\color[HTML]{000000} \textbf{marital-status}} & {\color[HTML]{000000} \textbf{occupation}} & {\color[HTML]{000000} \textbf{relationship}} & {\color[HTML]{000000} \textbf{race}}                 & {\color[HTML]{000000} \textbf{sex}}                   & \cellcolor[HTML]{FFCB2F}{\color[HTML]{000000} \textbf{\begin{tabular}[c]{@{}c@{}}capital-\\ gain\end{tabular}}} & {\color[HTML]{000000} \textbf{\begin{tabular}[c]{@{}c@{}}capital-\\ loss\end{tabular}}} & {\color[HTML]{000000} \textbf{\begin{tabular}[c]{@{}c@{}}hours-per-\\ week\end{tabular}}} \\ \midrule
{\color[HTML]{000000} 59}           & {\color[HTML]{000000} Self-emp-not-inc}   & {\color[HTML]{000000} Married-civ-spouse}      & {\color[HTML]{000000} Sales}               & {\color[HTML]{000000} Husband}               & {\color[HTML]{000000} White}                         & {\color[HTML]{000000} Male}                           & {\color[HTML]{000000} 0}                                                                                        & {\color[HTML]{000000} 0}                                                                & {\color[HTML]{000000} 40}                                                                 \\
{\color[HTML]{000000} 31}           & {\color[HTML]{000000} Private}            & {\color[HTML]{000000} Married-civ-spouse}      & {\color[HTML]{000000} Sales}               & {\color[HTML]{000000} Husband}               & {\color[HTML]{000000} White}                         & {\color[HTML]{000000} Male}                           & {\color[HTML]{000000} 0}                                                                                        & {\color[HTML]{000000} 0}                                                                & {\color[HTML]{000000} 40}                                                                 \\
{\color[HTML]{000000} 45}           & {\color[HTML]{000000} Self-emp-not-inc}   & {\color[HTML]{000000} Married-civ-spouse}      & {\color[HTML]{000000} Prof-specialty}      & {\color[HTML]{000000} Husband}               & {\color[HTML]{000000} White}                         & {\color[HTML]{000000} Male}                           & {\color[HTML]{000000} 0}                                                                                        & {\color[HTML]{000000} 0}                                                                & {\color[HTML]{000000} 40}                                                                 \\
{\color[HTML]{000000} 63}           & {\color[HTML]{000000} Private}            & {\color[HTML]{000000} Separated}               & {\color[HTML]{000000} Prof-specialty}      & {\color[HTML]{000000} Not-in-family}         & {\color[HTML]{000000} White}                         & {\color[HTML]{000000} Female}                         & {\color[HTML]{000000} 0}                                                                                        & {\color[HTML]{000000} 0}                                                                & {\color[HTML]{000000} 40}                                                                 \\
{\color[HTML]{000000} 17}           & {\color[HTML]{000000} Private}            & {\color[HTML]{000000} Never-married}           & {\color[HTML]{000000} Prof-specialty}      & {\color[HTML]{000000} Own-child}             & {\color[HTML]{000000} White}                         & {\color[HTML]{000000} Male}                           & {\color[HTML]{000000} 0}                                                                                        & {\color[HTML]{000000} 0}                                                                & {\color[HTML]{000000} 40}                                                                 \\
{\color[HTML]{000000} 42}           & {\color[HTML]{000000} Private}            & {\color[HTML]{000000} Divorced}                & {\color[HTML]{000000} Adm-clerical}        & {\color[HTML]{000000} Not-in-family}         & \cellcolor[HTML]{FE0000}{\color[HTML]{000000} Black} & \cellcolor[HTML]{FE0000}{\color[HTML]{000000} Female} & {\color[HTML]{000000} 0}                                                                                        & {\color[HTML]{000000} 0}                                                                & {\color[HTML]{000000} 38}                                                                 \\ \bottomrule
\end{tabular}
\end{table*}

\begin{figure*}[t]
\begin{center}
\includegraphics[width=\linewidth,height=6cm]{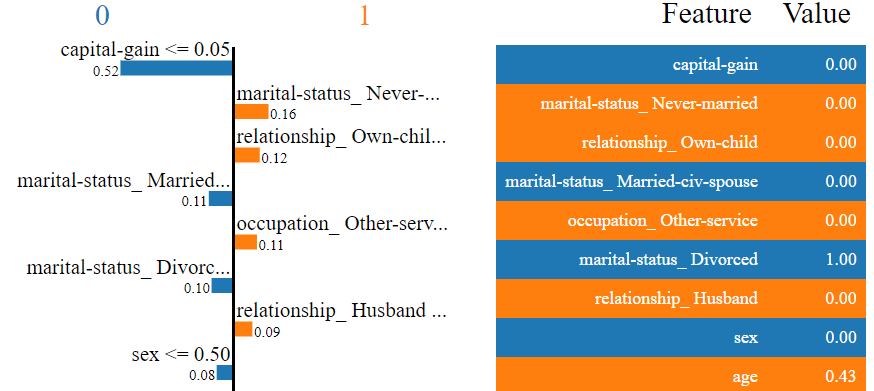}
\end{center}
\caption{LIME\cite{10.1145/2939672.2939778} showing the feature importance for the same data point (last row of Table 1).}
\label{LIME}
\end{figure*}

\begin{figure*}[t]
\begin{center}
\includegraphics[width=\linewidth]{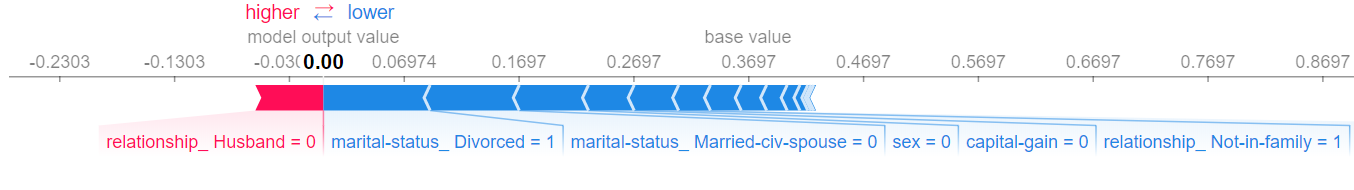}
\end{center}
\caption{SHAP\cite{NIPS2017_7062} showing the feature importance for the same data point (last row of Table 1).}
\label{SHAP}
\end{figure*}

\section{Dataset}
This paper is limited to binary classification models and tabular data (row-column format). We used a real-world, publicly available dataset - ``Adult Census Income'' which is widely used in \textit{Software Fairness} domain\cite{Angell:2018:TAT:3236024.3264590,Udeshi_2018}. This dataset consists of survey results of the income of 48,842 adults all over the USA\cite{ADULT}. The prediction task is to determine whether a person makes over 50K a year. There are two protected attributes - sex \& race.

\section{Proposed Framework}

Here we propose a k-nearest neighbor based explanation method to find bias in a classification model. In our experiment we use logistic regression model. At first, we pre-process the data. For categorical variables, one-hot encoding is used and numeric columns are discretized. In the ``Adult'' dataset, there are two protected attributes - sex (Male is privileged, Female is unprivileged)  \& race (White is privileged, Non-white is unprivileged). The last step of pre-processing is normalization. Then we divide the dataset into two parts - training (80\%) \& testing (20\%) with random shuffling.  After train-test division, we train the logistic regression model on the train data and test it on the test data. For any test data point, if the logistic regression model predicts negative outcome (Income is <=50K per year), we then find \textbf{K-nearest neighbor} data points (nearest to the point being tested) with positive class labels in the training data. Here K is a hyperparameter that can be tuned. We tried with different values such as 5,10,20. Once we get the K-nearest neighbor data points, we see the distribution of those points and compare it with the test data point. If they only differ on protected attributes (sex,race), then we say this data point is unfairly predicted. We raise an alarm. After rerunning the experiments ten times with random shuffling, we say on average ~10-17\% test data points are unfairly treated for this dataset by logistic regression model.

Table \ref{Adult} shows an example. The last row is the test data point which has been predicted to have negative outcome by logistic regression model. First five rows are similar data points having positive labels from the training data based on k-nearest neighbor approach. The most important features in this dataset are  ``capital-gain'' and ``capital-loss'' which are same for all the rows. So, that is not the deciding factor here. For validating, we perturb all the features to see whether prediction changes or not for the last row. When we change the protected attributes (Female to Male \& Black to White), the prediction changes to a positive outcome. Thus it is evident that the logistic regression model is discriminating this data point because of the value of protected attributes. This finding is similar with Galhotra et al.\cite{Galhotra_2017}. They found that in this ``Adult'' dataset, 11\% of the time,  prediction changes when the value of protected attribute is flipped.

We generate this tabular explanation for all test data points which are unfairly treated. A domain expert can easily evaluate our explanations and take decision whether to change the prediction or not. If the expert decides to change the outcome and assign new outcome by majority voting of neighbors, it is a post-processing approach (please see section \ref{Background}) of bias mitigation. Thus our approach finds the data points which are unfairly treated, gives explanation of the unfairness and also gives a choice of changing the outcome to make the prediction fair. Unlike LIME, SHAP or other currently used explanation methods we do not generate samples randomly. Instead we find similar data points from training data. Thus our explanation is more reliable when it comes to find bias.

We compared our explanation with LIME and SHAP. Figure \ref{LIME} and Figure \ref{SHAP} show the explanation provided by LIME and SHAP respectively for the same data point (last row of Table 1). For LIME (Fig \ref{LIME}), the features in ``blue'' color are responsible for negative outcome and features in ``orange'' color are responsible for positive outcome. The feature importance chart is showing ``capital-gain'' is the most important feature which is globally true but not important here as all the nearest data points have the same value of ``capital-gain''. LIME has ranked ``sex'' as the eighth important feature and does not report ``race'' at all. Seeing this explanation, it is very hard to conclude that this data point is unfairly treated. For SHAP (Fig \ref{SHAP}), the features in ``blue'' color are responsible for negative outcome and features in ``red'' color are responsible for positive outcome. SHAP explanation is showing that ``marital-status'' is the feature which is pushing this data point towards the negative outcome. SHAP is complaining about one protected attribute (sex) but is not reporting anything about the other protected attribute (race). Thus SHAP explanation also does not complain about unfair prediction for this particular data point. It is evident that our explanation is more valid in this scenario.


\section{Future Direction}
We have described a preliminary approach for finding, explaining and mitigating model bias and validated our approach for logistic regression model on a widely used dataset. Our immediate next step will be experimenting with different models and datasets. After that, we have a plan to create a tool for professional usage. We are summarizing our future course of actions to reach that goal:
\bi

\item In this paper, we are doing explanation based on singular instance (a single data point). But for any organization, making a future plan or creating policies is very difficult based on instance-based explanation. So, our future idea is to do global explanation. That will help an organization to make policies. We will repeat instance-based explanation for all the test data points and summarize the findings. Instead of just showing how an individual can get fair outcome, we will suggest a list of actions (plan) to give fair outcome to every individual. Thus we can combine fairness, explanation and planning. 

\item Future work will explore different K-nn algorithms such as ``ball tree'' and distance metrics such as ``manhattan'', ``chebyshev'', ``minkowski'', ``wminkowski''. These are hyperparameters which we will optimize based on the application. 

\item In this work while measuring distance every feature has been given equal importance. Depending upon the application, domain experts can suggest which attributes are more important than other. Important attributes will have higher power while measuring the distance. Thus instead of using default distance metric we need to generate a customized distance metric. 

\item Instance-based explanation is related with \textbf{individual fairness}. If the same approach is repeated for all the test data points but group wise (such as Male \& Female), then it will be applicable for \textbf{group fairness} too.

\item One limitation of our approach is the assumption that test data does not have a completely different distribution from training data. If that is the case, then the data points we will find using K-nn as nearest neighbors will not provide an accurate explanation for the discrimination. In that situation, we need to randomly generate samples around the instance by perturbing each feature like LIME. We should mention that this explanation could be unreliable. This is an interesting problem which will motivate future researchers to dig deep.

\ei

\section{Conclusion}
This NIER paper describes the shortcomings of current fairness measures and problems of finding underlying ML model bias using current explanation methods. A metric-free, K-nn based preliminary method is then proposed to explain \& visualize bias in the model behavior. Initial results of this approach outperform two state of the art explanation tools - LIME\cite{10.1145/2939672.2939778} \& SHAP\cite{NIPS2017_7062}.

Though software engineering community has taken the fairness problem of ML software seriously, we see very few papers are coming out from SE researchers. We hope this paper will motivate current SE community to actively work on this domain and collaborate with ML researchers \& industries to solve this problem together.

\balance
\bibliographystyle{ACM-Reference-Format}
\bibliography{main}

\end{document}